# Extreme Ultrafast Dynamics of the Refractive Index in Transparent Conductive Oxides: Theory and Experiment

**Noa Konforty[1,2†], Graham G. Brown[3†], Ohad Segal[2,4†], Oded Schiller[2,4], Yonatan Plotnik[1,2], Vladimir M. Shalaev[6], Misha Ivanov[2,3,5] and Mordechai Segev[1,2,4]***

*[1]Physics Department, Technion - Israel Institute of Technology; Haifa, Israel*

*[2]Solid-State Institute, Technion - Israel Institute of Technology; Haifa, Israel.*

*[3]Max Born Institute, Berlin, Germany*

*[4]Electrical and Computer Engineering Department, Technion; Haifa, Israel*

*[5]Institut für Physik, Humboldt-Universität zu Berlin, Berlin, Germany*

*[6]School of Electrical and Computer Engineering, Birck Nanotechnology Center and Purdue Quantum Science and Engineering Institute, Purdue University, West Lafayette, IN 47907, USA*

*† These authors contributed equally to this work*

*Corresponding author. Email: msegev@technion.ac.il*

## *Abstract*

Recent experiments in transparent conductive oxides (TCOs) have revealed light-induced order-unity variations in the refractive index occurring at extreme time scales, as short as a few-femtoseconds. These experimental observations remain unexplained, especially the ultrafast 10-20 femtoseconds relaxation of the index change, that cannot be explained by known phonon-mediated relaxation processes. Here, we present a simplified model followed by comprehensive simulations describing the phenomena, relying on the microscopic dynamics of electrons in TCOs under powerful ultrafast laser pulses. With this physical model, we predict and experimentally observe the unexplored regime of intraband modulation of electrons in the conduction band, leading to ultrafast oscillations of the refractive index. The observation of the oscillations validates the theory as a predictive tool, utilizing it to design experiments targeting novel effects that hinge on extreme alterations of optical properties of materials, such as photonic time-crystals and a plethora of novel extreme ultrafast phenomena.

## A. Introduction

The ongoing quest to achieve extreme temporal manipulation of the optical properties of materials is currently a major frontier in optics. Conceptually, changing the refractive index at few femtoseconds (fs) timescales by order unity enables the observation of fundamental phenomena such as time-reflection[1–5], ultrafast optical switching[6], generation of pairs of entangled photons[7], inverse prism[8] and temporal aiming[9]. Beyond a single modulation step, generating multiple temporal oscillations could unlock unprecedented regimes of time-varying optics. Among them the realization of photonic time-crystals[10–15] and their associated novel light-matter interactions[7,16–20], as well as other novel phenomena such as reciprocity breaking[21], extreme energy transfer[22] and time-domain bound states in the continuum[23,24].

These exciting phenomena seemed beyond reach for a long time. Over the years, various experiments have demonstrated considerable progress in the ultrafast modulation of the refractive index, by operating near the epsilon near zero (ENZ) point of transparent conductive oxides (TCOs)[25–33]. These pioneering observations were in the regime of ~100 fs modulations, where the interpretation of the underlying physics relied on the two-temperature model[27,34–38]. These large ultrafast changes in the refractive index enabled the observation of time-refraction[28,39] and various nonlinear processes at time-varying materials[40–43]. However, the time scale within which the index change was happening was far too slow for observing time-reflection and, more importantly - photonic time-crystals - which rely on the interference between multiple time-reflections and time-refractions[10–13]. The hope to observe these exotic phenomena came from more recent experiments where the variation in the refractive index occurs in the regime of a single optical cycle[44] and sub-cycle[5,45], enabling the measurement of time-reflection at optical frequencies[5]. In these experiments, the refractive index is varied by powerful laser pulses, which induce the excitation of electrons either within the conduction band or from the valence band to the conduction band (intraband or interband excitations, respectively). These excitations modify the plasma frequency of the TCO sample, causing a large change in the real part of the dielectric constant while keeping the change in the imaginary part small, allowing the material to remain relatively transparent. The change in plasma frequency in interband and intraband excitations arise from different physical mechanisms: interband excitation involves an increase in electron density, while intraband excitations causes a change in the plasma frequency due to the non-parabolicity of the conduction band[46–48].

The time scales within which the refractive index varies are of major importance, because they determine the ability to observe time-reflection and consequently photonic time-crystals. In this context, the excitation of electrons by light relies on absorption, which is an induced process and is therefore instantaneous. The relaxation back to the initial state is different. Conventionally, within the conduction band, the relaxation of the "hot" electrons was understood to occur via electron-phonon scattering, with timescales of hundreds of fs. However, recent experiments using intraband excitation revealed, unexpectedly, rapid relaxation – shorter than 20 fs[44]. This rapid decay directly contrasts with the anticipated hundreds-of-fs relaxation time of hot electrons in TCOs, which is governed by electron-phonon coupling. These observations challenge conventional theoretical frameworks, such as the two-temperature model[27,34–38,49], as they typically cannot account for the exceptionally fast relaxation times of the refractive index[50].

Driven by these anomalous observations, several recent theoretical efforts have sought to explain the complex refractive index dynamics and the physical origins of the ultrafast relaxation in TCOs. Notably, two recent works[51,52] propose alternative theoretical frameworks aimed at capturing the rapid order unity transient response, including the ultrafast relaxation. These works model the coherent electron dynamics under femtosecond modulation, revealing a mechanism that can, in principle, explain the ultrafast relaxation of electrons under intraband transitions induced by ultrashort laser pulses. However, despite these crucial steps forward, the field still lacks a clear physical model that not only resolves existing discrepancies but also accurately predicts new optical phenomena in extreme time-varying media.

Here, we propose a comprehensive, first-principles, theoretical framework that describes the microscopic dynamics of electrons within the valence and conduction bands of TCOs, and the fundamental mechanisms driving the extreme ultrafast refractive index variations observed in recent experiments. By modeling the dynamics of the electron density and explicitly accounting for the non-parabolicity of the conduction band, our framework captures the transient currents and the macroscopic polarization under powerful driving laser fields. Using indium tin oxide (ITO) as a case study, we show, through comprehensive simulations, that our model successfully reproduces the fast relaxation dynamics reported by Lustig et al.[44]. Furthermore, our model predicts a novel regime of modulation, where the electron distribution is pushed beyond the inflection point of the conduction band, causing the refractive index to exhibit complex temporal dynamics, where the refractive index oscillates with order-unity magnitude on the few-fs timescale. By using laser pulses that are even stronger than those used in Ref. 44, we observe this predicted dynamics experimentally in ITO[53]. These ultrafast oscillations of the refractive index change were first presented at the CLEO 2025 conference[53], and have recently been the focus of research by several groups [54–56], offering different models. Finally, we harness these insights to propose new strategies for optimizing the properties of the few-fs modulation pulses and for designing next-generation materials for extreme light manipulation, making an important step towards enabling the generation of photonic time-crystals.

## B. Toy model: intraband modulation of the effective mass

To understand the main physical mechanism responsible for the ultrafast refractive-index dynamics, we first consider a simple intraband excitation model. The purpose of this model is not to reproduce the full dynamics in ITO, but to isolate the key effects that enable the ultrafast modulation of the refractive index and generation of PTCs. Specifically, a strong optical pulse (henceforth "modulator") displaces, on the few-fs timescale, conduction-band electrons in a nonparabolic band, thereby changing the instantaneous band-curvature experienced by the electrons. Since the intraband optical response is determined by the plasma frequency, which is directly affected by the band curvature (the effective mass of the electron), the modulator pulse changes the Drude model parameters and therefore the refractive index.

For simplicity, consider a two-band material with a large bandgap compared with the photon energy of the modulator pulse, such that the modulator pulse does not change the overall density of conduction electrons. As in most recent experiments with light in time-varying media, the linear response of the material is probed by another (weak) optical pulse, henceforth the "probe", at a wavelength in the vicinity of the ENZ wavelength. The condition for ENZ is established by doping (giving rise to electrons at the bottom of the conduction band) such that the linear response of the conduction electrons compensates the linear response of the valence band electrons at the wavelength of the probe, resulting in a very small real part of the dielectric constant. Possible interband excitation (by multi-photon absorption of the modulator pulse) is initially neglected, hence the response is dominated by non-resonant interband polarization and the optically-induced intraband current. This approximation is appropriate for isolating the intraband response and is relaxed in the full numerical simulations (Sections C and E), where we show that the model holds predictive power despite its assumptions, and the dominance of the interband excitations can be controlled through selection of the system parameters.

Consider the vector potentials of modulator and probe pulses propagating along z and polarized along $x$, sampled at a single point in space:

(1)
$$A_{probe}(t) = A_p \cdot f_p(t) \sin(\omega_p t)$$
$$A_{mod}(t) = A_m \cdot f_m(t) \sin(\omega_m t)$$

Here $A_{p\backslash m}$ , $f_{p\backslash m}(t)$ and $\omega_{p\backslash m}$ are the vector potential amplitudes, normalized temporal envelopes and frequencies of the probe and modulator, respectively. The strong vector potential of the modulator pulse shifts the crystal momentum according to the acceleration theorem[57]: $k_x(t) = k_x + A_m f_m(t)\sin(\omega_m t)$ . In this section we use atomic (Hartree) units ($\hbar = e = 1$).

We approximate the initial distribution of conduction electrons (originating from doping) as equally distributed in $k_x, k_y, k_z \in [-k_0, k_0]$ (see Fig. 1c), with density $\rho_0$. The total electron number is $n_e = \rho_0 \cdot (2k_0)^3$. As in all such nondegenerate experiments, $\omega_{probe} < \omega_{pump}$, the modulator and probe pulses are not phase-locked. Therefore, we average over the modulator cycle when we consider the linear response to the probe pulse. The modulator-cycle-averaged driven system creates an effective dressed system sensed by the probe pulse. When the modulator and probe pulses overlap, the probe experiences an effective electron density $\rho_{eff}(k,t)$, which we calculate as the mean electron density, averaged over one cycle of the modulator pulse, centered symmetrically around time t. The remaining time-dependence follows the modulator envelope

$$\rho_{eff}(k,t) = \frac{\omega_p}{2\pi}\int_{t-\frac{\pi}{\omega_p}}^{t+\frac{\pi}{\omega_p}} \rho_{initial}\left(k_x + A_{mod}(t'), k_y, k_z\right)dt' \tag{2}$$

With the new effective distribution of the conduction electrons, the linear intraband response, to a weak modulator pulse polarized in the $x$ direction is:

$$\chi_c = -\frac{n_e}{\omega_p^2}\frac{1}{2k_0}\int_{BZ} dk_x\, \rho_{eff}(k_x)\frac{1}{m_e(k_x)} \tag{3}$$

where $m_e(k_x)$ is the effective mass of the conduction electrons at momentum $k_x$. Next, we find the full dielectric response of an ENZ material around the ENZ frequency. From our derivation of the intraband response, we obtain the final result:

(4) $$\varepsilon_{ENZ} = \varepsilon_{undoped}(\omega) - \varepsilon_{undoped}(\omega_0)\frac{\omega_0^2}{\omega^2}R\big(A_{mod}(t)\big)$$

$$R\big(A_{mod}(t)\big) = m_{eff,0}\frac{\omega_p}{2\pi}\int_{t-\frac{\pi}{\omega_p}}^{t+\frac{\pi}{\omega_p}} dt' \ \frac{\partial^2 \epsilon\big(A_{mod}(t')\big)}{\partial k_x^2}$$

Here, $\varepsilon_{undoped}(\omega)$ is the permittivity of the undoped material, $m_{eff,0}$ is the unmodulated averaged effective mass of the conduction electrons, $\omega_0$ is the ENZ frequency, and the last equality Eq.(4) is derived assuming $A_m \gg k_0$. Equation (4) is the central expression of the model. It shows that the modulator pulse modifies the intraband response (at the probe frequency) by changing the mean effective mass (band curvature) experienced by the conduction electrons. From Eq. (4), we can deduce the following conclusions:

First, in our model, the main contribution to the change in the refractive index is due to the change in the momentum distribution of the conduction electrons induced by the modulator pulse. This change is effectively instantaneous with the modulator pulse: it follows the modulator pulse both in its rise and in its relaxation. When the averaged effective mass increases, the response of the conduction electrons is suppressed, leading to rapid rise of the dielectric constant towards the value it would have had in the absence of doping. When the conduction electrons return to their original momentum, following the decrease in the modulator field, the dielectric constant decreases. This corresponds directly to the ultrafast rise and relaxation of the refractive index observed by Lustig et al [44].

Second, Eq. (4) presents an analytic approximation to the refractive index, that can be easily calculated given the lattice band-structure. To estimate the significance of this effect, we consider the simple tight-binding dispersion relation for the conduction band,

(5) $$u(k) = u_c - 2t_c\cos(ka_0)$$

Here $a_0$ is the lattice constant, $u_c$ is the on-site energy and $t_c$ is the nearest-neighbor hopping parameter. Explicitly evaluating Eq.(4a), we find that the maximal change in $\varepsilon$ becomes:

$$R(A_m) = J_0(A_m a_0) \quad (6)$$

$$\varepsilon_{ENZ}(\omega) = \varepsilon_{undoped}(\omega) - \varepsilon_{undoped}(\omega_0)\frac{\omega_0^2}{\omega^2}J_0(A_m a_0)$$

where $J_0$ is the zero-order Bessel function. For a lattice constant of about 1nm, as in ITO, and a modulator pulse at 800 nm, $A_m a_0$ approaches unity at intensity I=3.10$^{11}$W/cm$^2$, signifying very large and nearly instantaneous changes in the refractive index, which reaches 50% of its undoped value.

This model explains the observations by Lustig et al. [44] where the relaxation of the index change (induced by the modulator) occurs within 10-20 fs, similar to the rise time, which was thus far unexplained, and stood in direct contrast to the known explanations and predictions [27,34–38,48].

We now illustrate this model by applying Eqs. (3) and (4) to the parameters of the first conduction band of ITO in the $\Gamma P$ direction, Fig.1(a). The full bandstructure was obtained using the *QuantumESPRESSO* code[58,59], followed by the Wannier90 procedure[60] to generate tight-binding representation of the system Hamiltonian and thus obtain phase-consistent Berry connections for all bands included in the detailed simulations described in the following section, using the method of Ref.[61]. Figure 1c shows the effective electron distribution at different times for a given modulator pulse, and Fig. 1d shows the change in the dielectric function as a function of time, at a single frequency near the ENZ. Rapid increase in the effective mass of the conduction electrons, significant already at $k \sim \pm 0.05$, implies the increase in the dielectric function on the scale of unity for the frequencies in the vicinity of the ENZ point. We note that in modulator-probe experiments, the observables are usually the convolution of the probe pulse with the dielectric function, which may smooth out sharp features presented in this calculation.

Next, using the same conduction band, we consider a more exotic case occurring at higher intensities of the modulator pulse. As shown by the dispersion curve (Fig. 1a), away from the bottom of the conduction band there are several inflection points, with the first one being around $k = \pm 0.1$. Our model suggests that, driving the distribution of the conduction electrons past this point results in a change in the trend of the dielectric function. Namely, as shown in Fig. 1f, using a stronger modulator pulse yields oscillatory behavior of the dielectric function, all within a single modulator pulse at timescales faster than its temporal envelope. This new mechanism further increases the ability of ultrafast pulses to change the refractive index dramatically, even multiple times within a single modulator pulse.

To conclude this section, we note that the toy model can be extended to include interband excitations via multi-photon transitions, which are included in the comprehensive simulations described in the following sections.

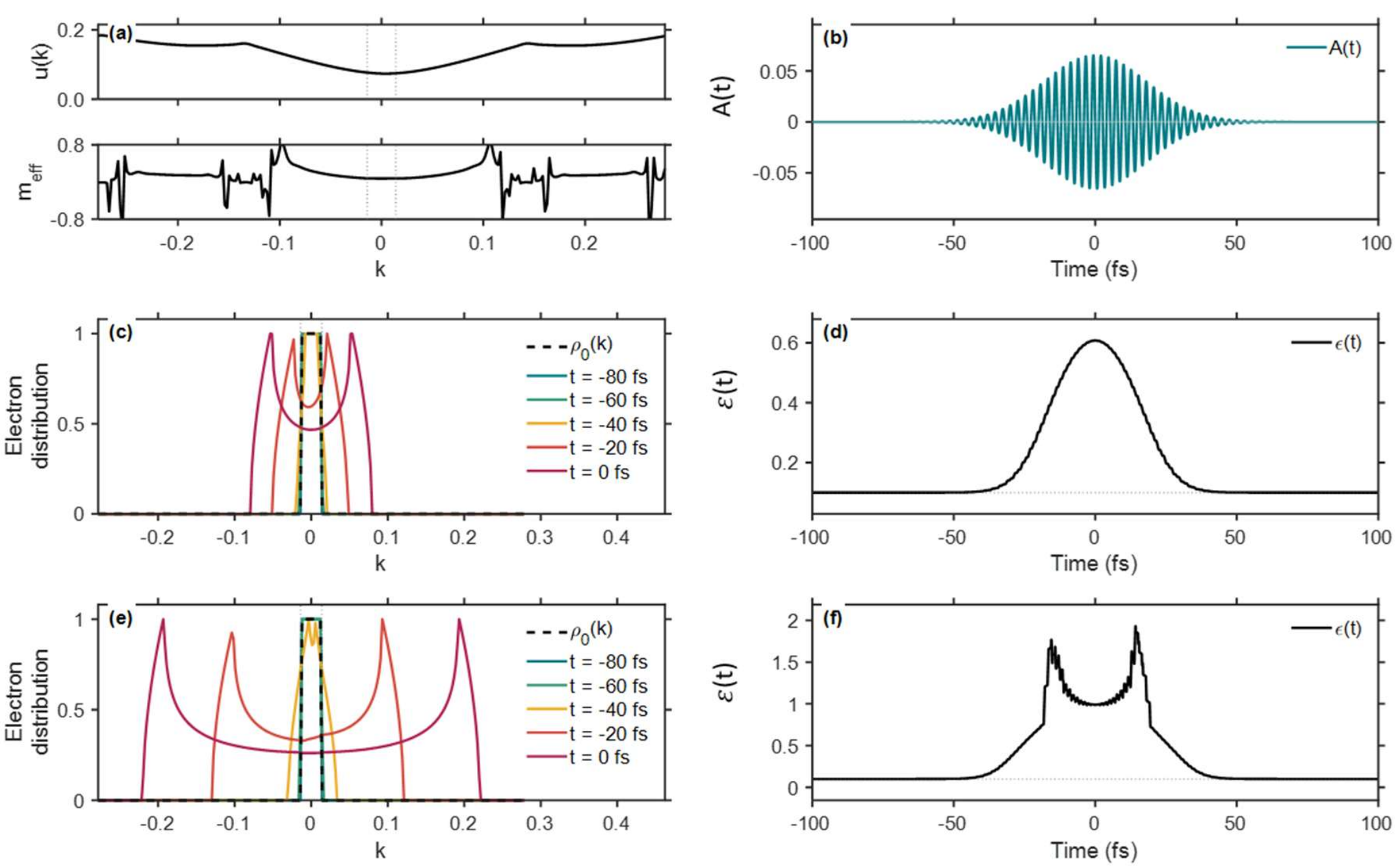

Figure 1: **Results calculated based on the toy model.** (a) Lowest conduction band of ITO in the $\Gamma - P$ direction (upper panel), and the electron effective mass, which is proportional to the inverse of the second derivative of the conduction band (lower panel). The first inflection points of the conduction band occur at $k\sim \pm 0.1$. (b) Vector potential of the modulator pulse, $A(t)$. (c) Normalized effective momentum distribution of conduction electrons with and without the modulator pulse, at different times. In the absence of the modulator pulse, the electrons are uniformly distributed in a rectangular region at the bottom of the conduction band (dashed). The modulator pulse pushes the electrons to higher momenta, in a fully reciprocal manner. (d) Dielectric response near the ENZ point as a function of time, for a given modulator pulse. The dielectric response rises and relaxes back to its original value, following the envelope of the modulator pulse, explaining the ultrafast relaxation of the refractive index. (e)-(f) Same as (c)-(d) respectively, with a stronger modulator pulse. By increasing the peak intensity of the modulator pulse, the effective electron distribution reaches farther in the momentum space, crossing the inflection point and resulting in the oscillations of the dielectric function $\varepsilon(t)$ shown in (f).

## C. Full numerical simulations in ITO

To validate the conclusions of our toy model and gain further insight, we model the microscopic dynamics of conduction electrons in ITO. We use the density functional theory to create a tight-binding Hamiltonian of ITO (described in the previous section) and solve the semiconductor Bloch equations within the single active electron approximation. Initially, the system is doped in the first conduction band to achieve ENZ conditions at the desired probe wavelength. With this initial population of conduction electrons, we simulate the evolving dynamics under a strong modulator pulse driving the electronic system out of equilibrium, and calculate the linear response of the medium for the 60fs probe pulse centered at the ENZ frequency as a function of modulator-probe delay, $\tau$. From the linear response, we obtain the reflection and transmission coefficients for a thin ITO film, which enables the comparison with our experimental results. For more details on the simulations see Methods Section.

**C.1. Explanation of the observed few-fs rise and relaxation of the dielectric response**

Here, we aim to simulate the response of the medium under parameters similar to the experiment by Lustig et al[44]. Figure 2a shows the reduced band-structure of ITO used in the computation. The results demonstrate the ultrafast rise of the refractive index followed by the ultrafast relaxation, with a weak long tail. The fast rise and relaxation of the refractive index resemble those observed in the experiments – in magnitude, sign, and their rise and fall times. In fact, the fast rise and the fast relaxation of the index change are almost instantaneous with the modulator pulse. These ultrafast changes are smoothed over by the cross-correlation of the modulator and the probe. The simulations confirm that the main contribution to the index change comes from coherent intraband currents induced by the modulator pulse. The long tail in the relaxation is indicative of additional, less dominant, processes that have slower relaxation rates. Numerically, they arise from the relaxation of the relatively small number of valence electrons photo-injected into the first conduction band. The relaxation mechanisms include intraband relaxation due to coupling to phonons, with a time constant set at 100fs, and interband relaxation at a much slower, picosecond, time-scale.

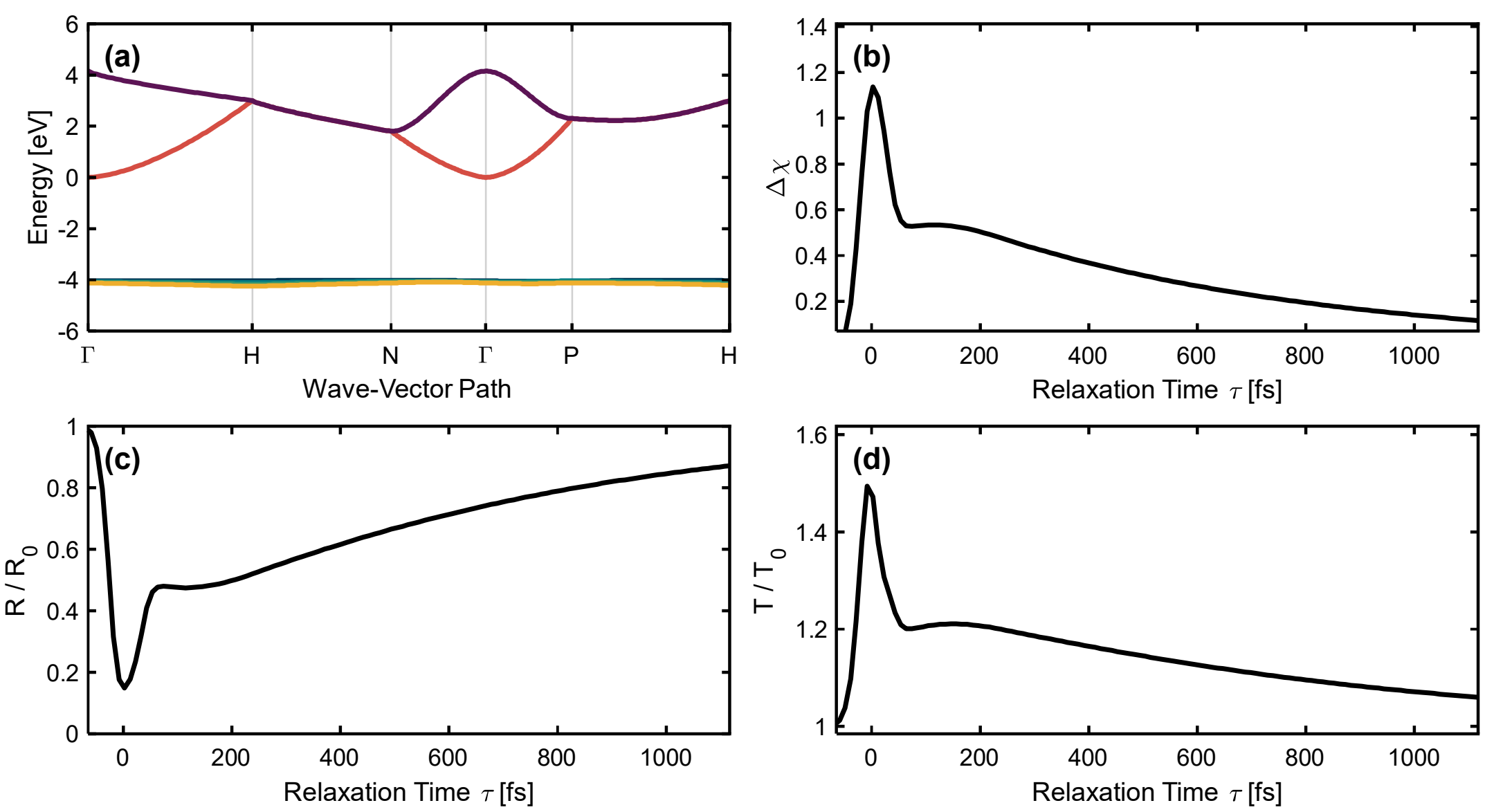

Figure 2: **Simulations in ITO showing ultrafast relaxation of the refractive index**. **(a)** Reduced band-structure of ITO, obtained using *QuantumESPRESSO*. **(b)** Variation in the (averaged) real part of the susceptibility experienced by the probe, as a function of modulator-probe delay, for a $2\text{TW/cm}^2$ , 6fs, modulator pulse. The susceptibility rises and relaxes following the envelope of the modulator pulse. **(c)-(d)** Normalized transmission **(c)** and reflection **(d)** coefficients calculated from susceptibility experienced by the probe, for each time delay.

## C.2. Prediction of oscillations in the dielectric response

Next, we explore the dynamics under an ultra-strong modulator pulse that drive the electrons past the inflection point in the conduction band. In this regime, the simulations show that, when the peak intensity of the modulator pulse is sufficiently high, the dielectric response and the transmission and reflection coefficients oscillate above and below their original value, within the duration of each modulator pulse (Fig. 3).

It is important to note that, in the full simulations, the electron dynamics include decoherence, thermalization, interband excitations, and electron–electron scattering, in addition to the intraband currents described by the toy model. We find that the intraband contributions remain dominant, giving rise to the ultrafast oscillations. However, as the modulator pulses become more intense, interband effects become increasingly significant and contribute to the dynamics producing longer temporal tails, seen both in Fig. 2 and 3.

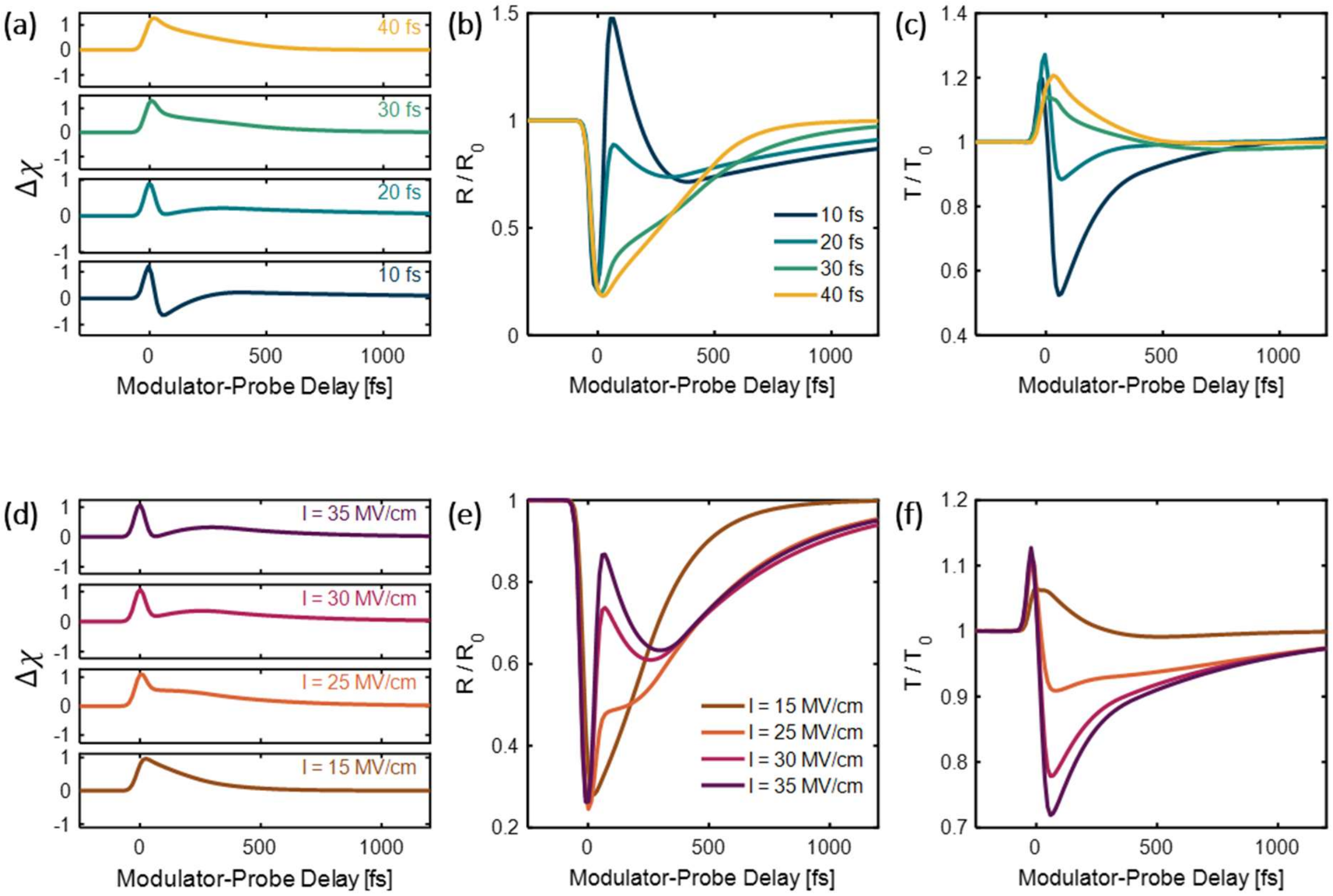

Figure 3: **Simulations in ITO predicting oscillations in the refractive index. (a)** Averaged change in the real part of the susceptibility as a function of modulator-probe delay for different durations of the modulator pulse at a constant pulse energy. As the modulator is compressed to shorter durations and the peak power increases, the susceptibility oscillates above and below its unperturbed value. $\chi$ **(b)-(c)** Normalized reflection **(b)** and transmission **(c)** calculated from the averaged susceptibility the probe experiences for each time delay. Both coefficients display the same oscillatory behavior. **(d)-(f)** Same as **(a)-(c)** for different modulator energies given in units of MV/cm, and fixed duration of 20 fs.

The relative strength of these mechanisms can be tuned by selecting a material with a larger bandgap or by using longer-wavelength modulator pulses. Under these conditions, the interband processes will be suppressed, while the ultrafast intraband dynamics is preserved. Nevertheless, even in the high-intensity regime of the experiment, the toy model retains its qualitative predictive power and agrees well with the results of the full simulations.

## D. **Experimental observation of oscillatory dynamics in the dielectric response**

To observe the oscillatory dynamics predicted by both the toy model and the full ITO simulations, we perform a modulator–probe experiment in thin ITO films. The experimental setup consists of two laser pulses: a modulator pulse centered at 800 nm and a probe pulse centered at 1200 nm. The modulator pulse is compressed to various pulse durations of ranging from 11 fs to 40 fs FWHM (full-width-at-half-maximum), maintaining the same pulse energy. This is accomplished using a hollow-core fiber system with varying gas pressure, chirped mirrors and fused silica slabs for adjusting the spectral phases. The total energy of the modulator pulse is controlled using a half-wave plate and a polarizer placed in front of the hollow-core fiber system. The duration of the modulator pulse is determined using FROG. The probe pulse has a duration of 60 fs FWHM and is generated using an optical parametric amplifier. The two pulses are made to spatially overlap on a 310 nm-thick ITO sample, judiciously doped to have an ENZ wavelength of 1225 nm. The relative delay between the modulator and probe pulses is controlled using a delay stage. To characterize the modulator-induced change in the dielectric response of ITO, we measure the transmitted and reflected probe amplitudes as a function of the modulator–probe delay, as shown in Fig. 4.

We observe that, as the peak intensity of the modulator pulse is increased, either by shortening the pulse duration or by increasing the total pulse energy, a complex temporal response of the ITO sample emerges. The Fresnel reflection coefficient initially decreases while the transmission coefficient increases. Subsequently, we observe that both trends are reversed: the Fresnel reflection rises above its initial value, while the transmission drops considerably below its initial value. At later delays, the transmission relaxes back toward its original value, while the reflection undergoes an additional reversal. This behavior indicates that the dielectric function first increases and then decreases during the modulation pulse, in agreement with our simulations. Importantly, the observed complex dynamics occurs on the ultrashort timescale of the modulator pulse, followed by a longer relaxation tail. These timescales are consistent with those previously reported by Lustig et al.[44] and with the theoretical model presented here.

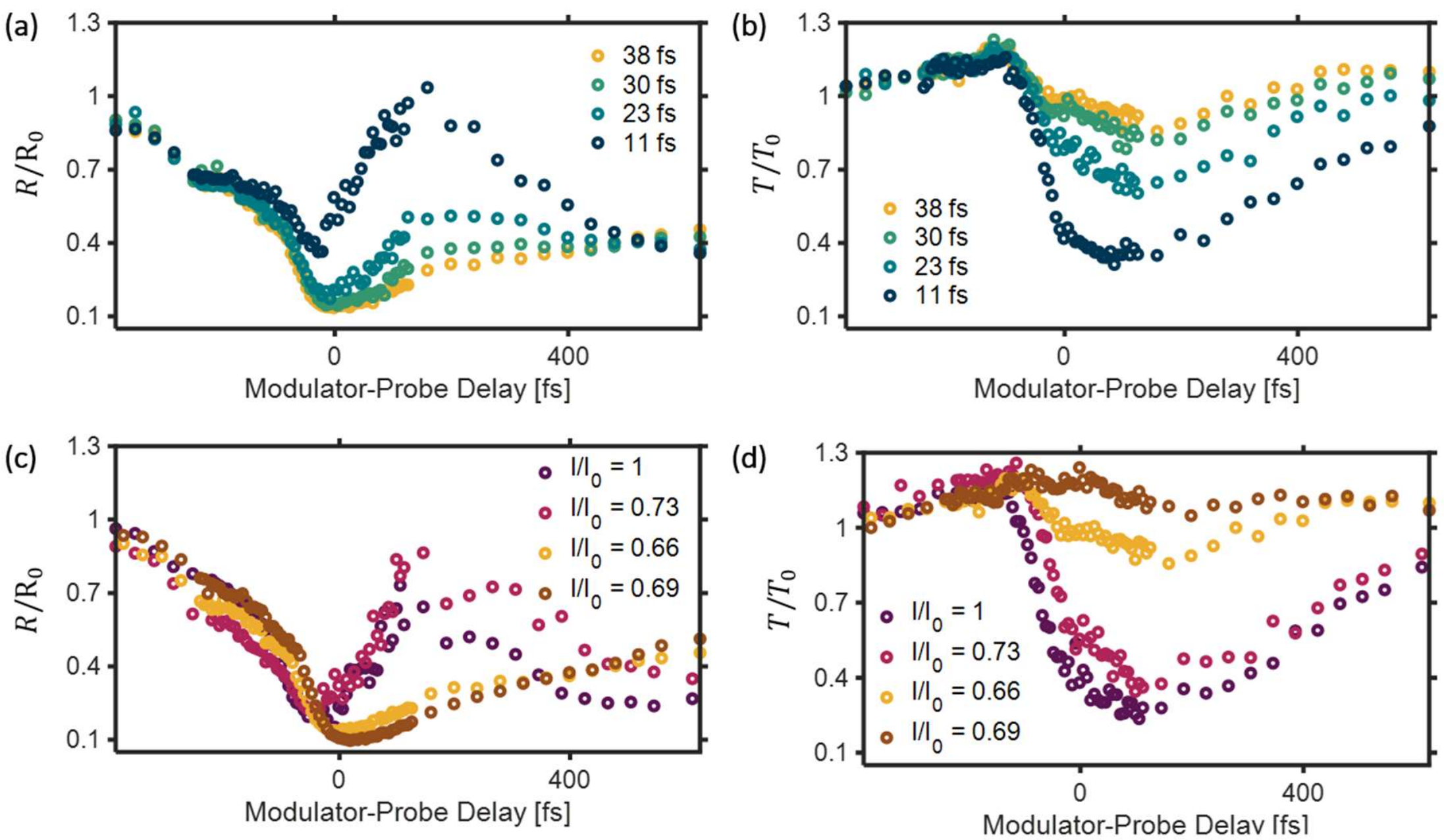

Figure 4: **Experimental results showing ultrafast complex dynamics of the dielectric response in ITO**: **(a-b)** Fresnel-reflected intensity of the probe pulse, normalized to the value of the unmodulated reflection **(a)**, and the respective transmitted intensity of the probe pulse **(b)**, as a function of modulator-probe delay. Different colors correspond to different durations (FWHM) of the modulator pulse. In all measurements the total modulator energy is kept fixed. **(c-d)** Same as in **(a-b)**, respectively, but for various modulator pulse energies. In all measurements the duration of the modulator pulse is fixed at 38 fs FWHM.

## E. Generation of PTCs

Building on the insights gained from the toy model and the simulation framework, we propose a route toward realizing PTCs. PTCs at optical frequencies require order-unity index variations occurring periodically in time, with few-fs period. With this consideration in mind, the slow relaxation of the refractive index, whether due to interband mechanisms or via coupling to phonons, would pose a severe limitation. The results presented above show that it is possible to exploit the extreme ultrafast relaxation of intraband excitations while suppressing the (much slower) interband dynamics. To achieve this under realistic conditions, we propose driving a carefully selected material with a train of ultrafast pulses. The material should have a sufficiently large bandgap, such that at least two- and three-photon absorption from the modulator pulse are suppressed. Our simulations suggest that this approach can produce periodic refractive-index modulation on a timescale of ~10-20fs. Figure 5 illustrates this concept for an ITO-like material, with the bandgap increased to 6eV and the dipole transition amplitudes scaled accordingly. The material is doped such that the ENZ is at $12\mu m$. The material is driven by an ultrashort pulse train at 800nm wavelength, giving rise to a PTC with 20fs period for a probe at $12\mu m$ wavelength. Our calculations show that materials combining ENZ response in the mid-infrared with a sufficiently large bandgap enable the realization of PTCs at optical frequencies.

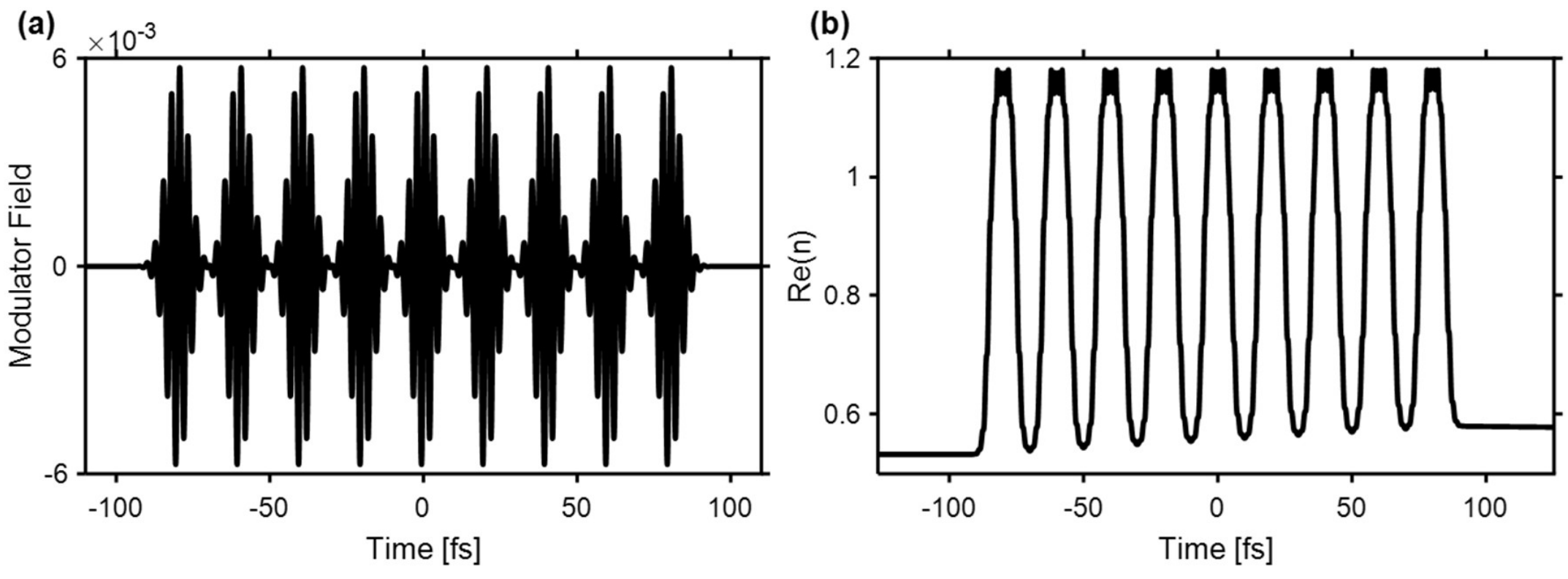

Figure 5: **PTC at optical frequencies in ITO with increased bandgap (6 eV), generated by consecutive modulator pulses at 800 nm wavelength. (a)** The modulator pulse train consisting of nine consecutive pulses. **(b)** Real part of the refractive index of the PTC at the ENZ point, calculated by applying Eq. 4 to the electron distribution in the conduction band resulting from the comprehensive simulations.

## F. Discussion

Following the initial observation[53] of the dynamic oscillations in the refractive index of ITO, several recent studies have proposed alternative explanations, generally attributing the dynamics to competition between intraband and interband excitation mechanisms [54–56]. In that interband mechanism, the excitation across the bandgap (whether via linear absorption of UV photons or through nonlinear multi-photon absorption) increases the density of conduction electrons, thereby increasing the plasma frequency and shifting the dispersion relation. As previously shown[6,62], such interband effect causes a decrease in the refractive index. An important difference between the intraband and the interband mechanism is the relaxation times. Namely, the intraband mechanism presented here has extremely short relaxation times, because the process relies on displacing electrons within the conduction band, effectively on the sub-cycle time-scale, with the displacement amplitude following the envelope of the modulator pulse. In contrast, for the interband mechanism, the relaxation involves recombination, which has typical times of hundreds of fs to picoseconds. Specifically for ITO, the relaxation time of interband excitations is on the order of 0.5 ps [63], while in our transmission measurements in Fig. 5 we see oscillations on the timescales of ~100-200 fs. As explained above, this difference is essential in planning experiments: for experimenting with isolated time-interfaces - the slow relaxation does not pose problems, whereas for studying the impact of sequential time-interfaces or for realizing PTCs, the slow relaxation is a showstopper.

Essentially, both the interband carrier injection and the intraband displacement beyond the inflection point can, in principle, explain the variations in the reflection and transmission observed in the experiments. Our simulations indicate that in ITO the intraband mechanism remains dominant until the modulator intensities reach ~ 3 TW/cm$^2$. Our simulations also show that excitation of valence electrons into the conduction band is accompanied by simultaneous de-excitation of the conduction-band electrons into the valence band, suppressing the net interband contribution. Last but not least, increasing the modulator wavelength reduces the interband excitation while increasing the strength of the refractive index variation driven by intraband carrier dynamics. Overall, the intraband carrier motion is the only mechanism capable of generating reversible ultrafast modulations of the refractive index occurring on the time scale of a few fs, as deduced from the modulator-probe cross-correlation measurements. Consequently, the ultrafast oscillations observed in the reflection and transmission reflect the displacement of the conduction electrons beyond the inflection point of the conduction band.

This issue is especially important in designing experiments to generate photonic time-crystals at optical frequencies, where the relatively slow relaxation following interband excitation could limit how short the modulation period could be. Our results show that it is possible to completely eliminate the interband contribution, removing the hurdle of the relatively slow relaxation. This can be done by using TCO materials with a wider bandgap, or using modulator pulses at longer wavelengths, thereby suppressing the interband injection exponentially. The intraband dynamics past the inflection point can be used to obtain modulation of the refractive index at twice the periodicity of the modulation train, further increasing the effect.

### G. Conclusion

We studied theoretically and experimentally the ultrafast optical response of ITO under strong-field modulation. We showed that the ultrafast relaxation of the dielectric response can be explained by coherent intraband motion of the conduction electrons, experiencing strong variations in their effective mass due to the non-parabolic nature of the conduction band. This leads to a shift in the (average) effective mass, manifested in a very large change in the refractive index. Once the conduction electrons are excited beyond the inflection point, this mechanism predicts oscillatory dynamics of the dielectric function during the modulator pulse. We observe these oscillations experimentally using modulator–probe measurements in ITO films near the ENZ wavelength, in agreement with the predicted dynamics.

Most importantly, our results show that ITO can support ultrafast reversible modulation of its optical response on the timescales shorter than the driving pulse envelope. Under periodic strong-field excitation, this response provides a route towards generating a photonic time-crystal in ITO, where the dielectric function is modulated periodically in time at sub-probe-cycle rates. These results suggest that intraband-driven transparent conducting oxides near their ENZ wavelength can serve as a platform for realizing strongly time-dependent media with sub-cycle variation at optical frequencies.

Acknowledgements: The Technion is supported by the Breakthrough Program (MAPATS) of the Israel Science Foundation (ISF) and the United States Air Force Office of Scientific Research (AFOSR).